\documentclass[8pt]{extarticle}

\usepackage[a4paper, margin={0.5in}]{geometry}

\usepackage[colorlinks,linkcolor=red]{hyperref}
\usepackage{amsmath}
\usepackage{amssymb}
\usepackage{caption}
\usepackage{subfig}
\usepackage{booktabs}
\usepackage{geometry}
\usepackage{pdflscape}
\usepackage{placeins}
\usepackage{graphicx}
\usepackage[utf8]{inputenc}
\usepackage[english]{babel}
\usepackage[parfill]{parskip}
\usepackage[numbers]{natbib}
\usepackage{url}

\begin{document}

\begin{center}
\textbf{Principal Process Analysis of dynamic GlucoCEST MRI data}\\
Stefano Casagranda\footnote{Olea Medical, La Ciotat, France},
Marco Pizzolato\footnote{EPFL, Lausanne, Switzerland},
Francisco Torrealdea\footnote{Centre for Medical Imaging, UCL, London, United Kingdom},
Xavier Golay\footnote{Institute of Neurology, UCL,
London, United Kingdom}, and
Timothé Boutelier\footnotemark[1]\\
\end{center}

\textbf{Synopsis.}
GlucoCEST is an MRI contrast enhancement technique sensitive to the concentration of sugar in the tissue. Because of a difference in metabolism, it
is thought that tumors consume more sugar than normal tissue. However, glucose metabolism is complex and depends on many processes, which
are all important to understand the origin of the measured signal. To achieve this goal we apply here a process analysis method to a deterministic
system describing the metabolism of glucose in the tissue.

\textbf{Introduction.} 
Chemical Exchange Saturation Transfer (CEST) is an MRI contrast enhancement technique that enables the indirect detection of molecules with exchangeable protons \cite{c1}. GlucoCEST is a CEST technique that measures a signal related to the concentration of injected glucose and its derivates \cite{c1}. It is expected that metabolic anomalies due to the presence of cancerous tissue could be measured or characterized by means of glucoCEST, and in particular to its dynamic characteristics. There the signal is analyzed as a function of time following glucose metabolism, which is paramount to pathological tissue assessment. Different subvoxel compartments contribute to the signal. Particularly, a simplified model made as a system composed by 19 Ordinary Differential Equations (ODEs) can be used in a first step, where different parameters can be set to simulate healthy or tumor-like conditions \cite{c2}. The system accounts for the presence of vascular glucose, its equilibrium within the interstitial environment, and the transport into the intracellular space. The purpose of this work is to identify the processes that contribute more to the signal. This analysis allows designing a reduced model, dedicated to the fitting of experimental data, and quantification of physiological parameters. To do so we apply Principal Process Analysis (PPA) \cite{c3,c4}, a numerical method for the analysis and reduction of biological systems designed with ODE formalism. PPA allows associating a dynamic weight to all involved processes and inferring their importance during the acquisition time.

\textbf{Methods.}
We consider each ODE of the system as a sum of biological processes. Let
\begin{equation}
\label{eq_1}
\dot{x}_i=\sum_j f_{i,j}(x,p)
\end{equation}
where $x=(x_1, x_2, \ldots,x_n) \in \Re^n$ is the vector of concentrations of glucose in different compartments and $p \in \Re^b$ is the vector of parameters. $f_{i,j}(x,p)$ is the $j^{th}$ process involved in the dynamical evolution of the $i^{th}$ variable of the system over the acquisition time $[0,T]$. A relative weight $W_{i,j} (t,p)$ is associated to each process $f_{i,j}(x(t),p)$, to study its influence in the evolution of the variable $x_i$ over time $t$:
\begin{equation}
\label{eq_2}
W_{i,j}(t,p)=\frac{|f_{i,j}(x(t),p)|}{\sum_j |f_{i,j}(x(t),p)|}
\end{equation}

where $0 \leq W_{i,j}(t,p)\leq 1$ and $\sum_j W_{i,j}(t,p)=1$. We use the thresholds $\delta$ and $\nu$ to detect not only the \textit{inactive} processes but also to make a further distinction between processes with a \textit{moderate activity} and a \textit{fully activity}.
We call a process $f_{i,j}(x(t),p)$ \textit{inactive} at time $t$ when $W_{i,j}(t,p) < \delta$,\textit{ moderately
active} at time $t$ when $\delta \leq W_{i,j}(t,p)< \nu$, \textit{fully active} at time $t$ when $W_{i,j}(t,p) \geq \nu$. Furthermore, because the glucoCEST signal $y$ is seen as a sum of the glucose concentration in different compartments (described by six variables of the model in \cite{c2}) we also compute the contribution of each variable $x_i(t)$ for the system output $y(t)= \sum_i x_i(t)$:
\begin{equation}
\label{eq_3}
M_i(t)=\frac{x_i(t)}{\sum_i x_i(t)}
\end{equation}
where $0\leq M_i(t) \leq 1$ and $\sum_i M_i(t)=1$. Their \textit{activity} is dictated in the same manner by the thresholds $\delta$ and $\nu$, set at 0.1 and 0.4. The outcome of the
analysis is summarized with an extended \textit{Boolean Process Map} \cite{c3} called \textit{3-Level Process Map} with three possible \textit{activity} outcomes.

\begin{figure}[h!]
\centering
\includegraphics[width=0.85\textwidth]{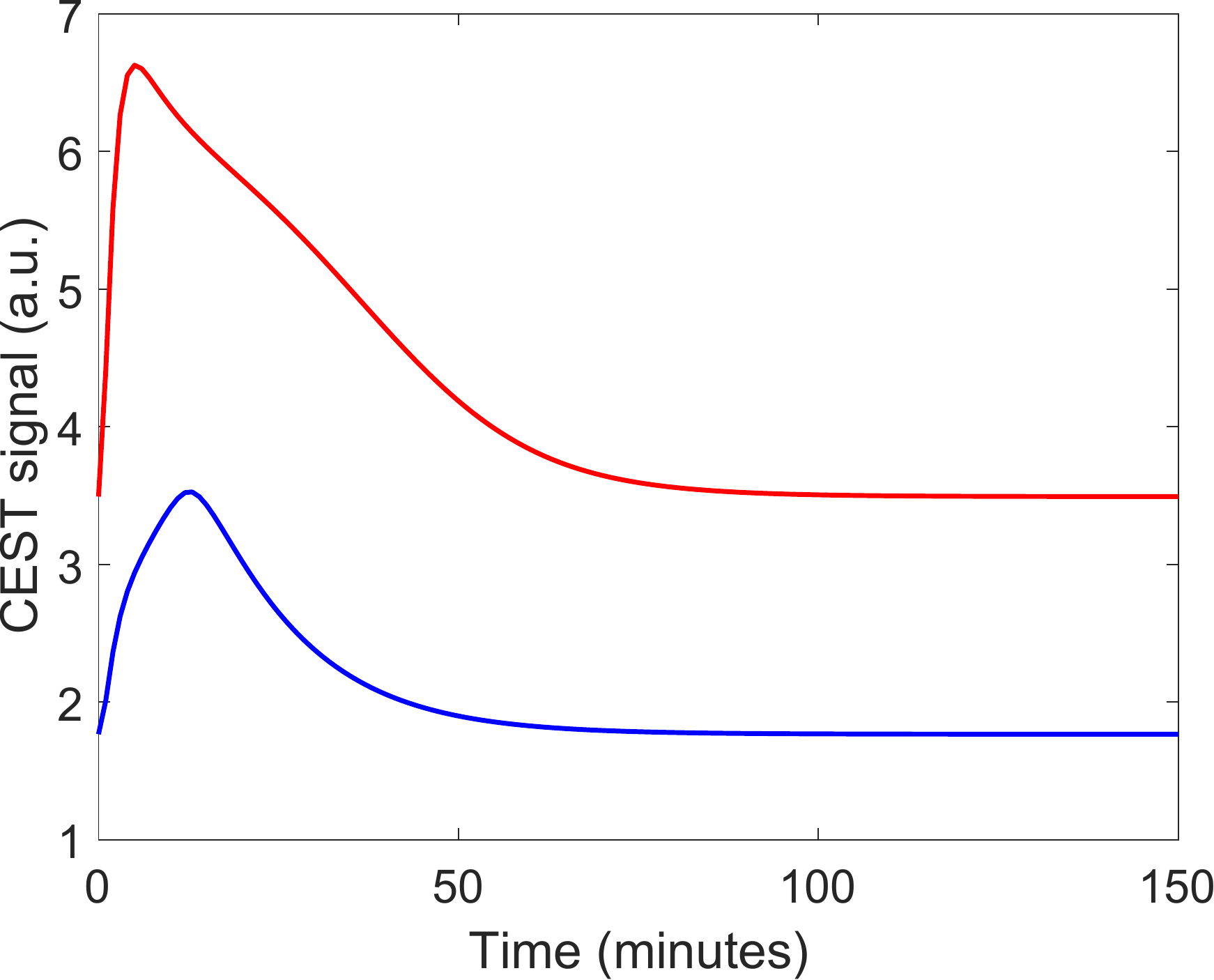}
\caption{CEST signal in a healthy tissue (blue color) and cancerous tissue (red color) during an acquisition time of 150 minutes.}
\label{fig_1}
\end{figure}

\begin{figure}[h!]
\centering
\includegraphics[width=0.85\textwidth]{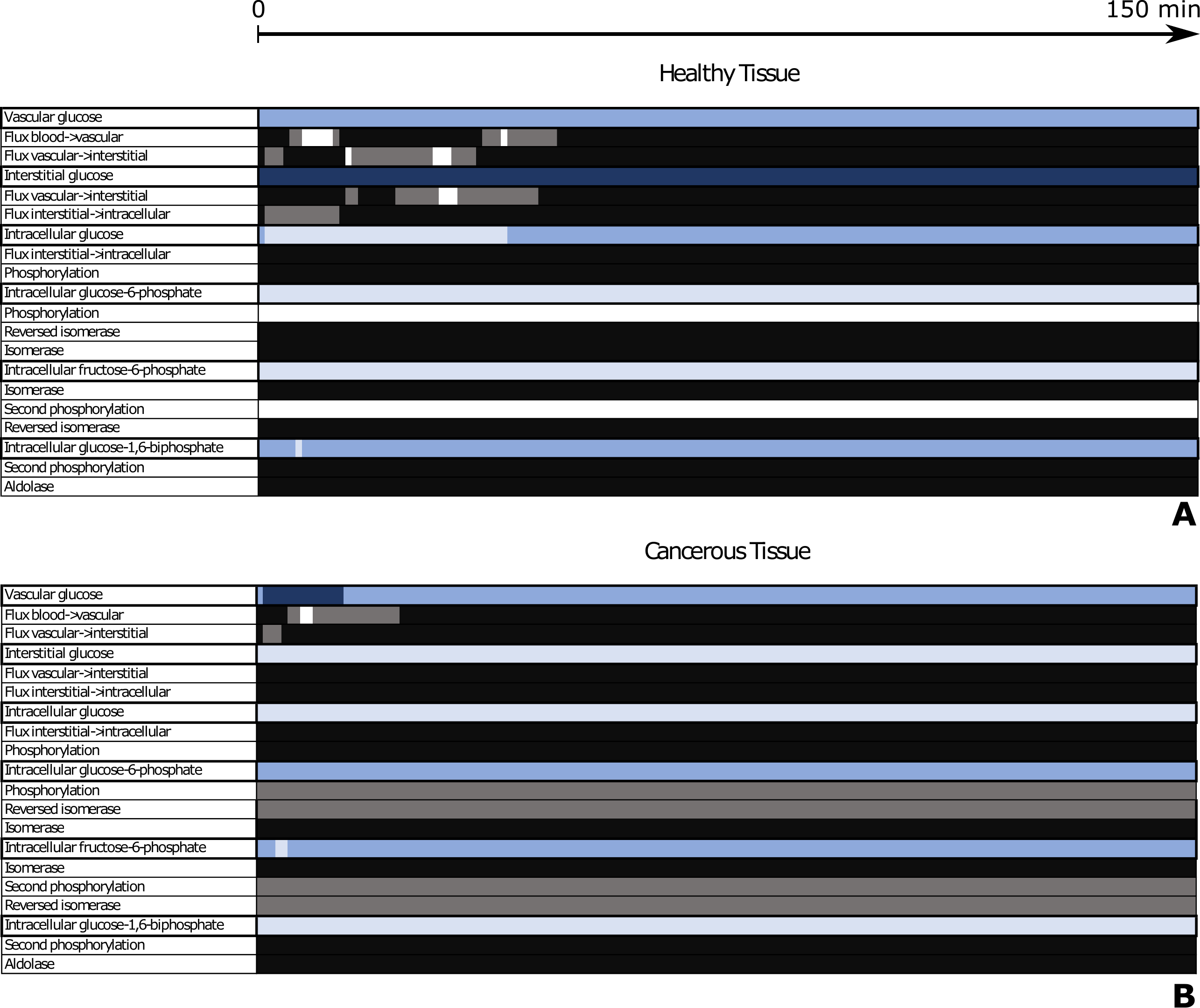}
\caption{\textit{Activity} of six model variables of glucoCEST model and of their processes during a 150-minute period in an healthy (A) and cancerous tissue (B). Processes are
listed in the first column (white background) ordered by variable (white background, in bold). Their \textit{activity} is depicted in the second column between 0 and 150
minutes. A horizontal dark blue (black) bar is displayed when the variable (process) is \textit{fully active}, blue (grey) when it is \textit{moderately active} and light blue (white)
when it is \textit{inactive}.}
\label{fig_2}
\end{figure}

\textbf{Results.}
Fig. \ref{fig_1} shows the two different CEST signals simulated in healthy and cancerous tissue after an IV bolus, using the parameters described in \cite{c2}. The two \textit{3-Level Process Maps} in Fig. \ref{fig_2} reports the \textit{activity} of the system variables that contribute to the CEST signal and of their relative processes, in both healthy and cancerous tissue.

\textbf{Discussion.}
The map in Fig. \ref{fig_2}A shows that in a healthy tissue the highest contribution in the generation of CEST signal is given by the glucose in the interstitial compartment as seen during an IP bolus situation \cite{c5}. Fig. \ref{fig_2}B shows that in the tumor case the interstitial contribution becomes \textit{inactive} because the glucose carriers bring the glucose faster into the cell. For the first minutes after the injection the vascular compartment is \textit{fully activ}e and then it becomes moderately active because most of the glucose bolus has passed through the tissue. From our theoretical developments, the intracellular compartment seems to be dominated by glucose and
fructose-1,6-biphosphate within healthy tissue (Fig. \ref{fig_2}A), while glucose-6-phosphate and fructose-6-phosphate dominates in cancerous tissue (Fig. \ref{fig_2}B).

\textbf{Conclusion.}
We have presented the PPA of a glucose metabolism model, in the context of dynamic glucoCEST imaging. It shows that dominant processes are different
depending on the properties of the tissue. It would be interesting to apply the technique for different kind of tumors and glucose administration. Based on this analysis we can build different reduced models where the processes that remain \textit{inactive} have been removed.

\textbf{Acknowledgments.}
This project has received funding from the European Union’s Horizon 2020 research and innovation programme under grant agreement No 667510 and the Department of Health’s NIHR-funded Biomedical Research Centre at University College London.

\textit{Reference to the original publication.}
In ISMRM 27th Annual Meeting.
Proc. Int. Soc. Mag. Reson. Med. 27 (2018).
%
%

%
\FloatBarrier
\hspace{20pt}
\begin{center}
\textbf{Figures}\\
\end{center}


\end{document}